\documentclass[preprints,article,accept,pdftex,moreauthors]{Definitions/mdpi} 
\firstpage{1} 
\pubvolume{1}
\issuenum{1}
\articlenumber{0}
\pubyear{2022}
\copyrightyear{2022}
\hreflink{https://doi.org/10.3390/universe8040232} 
\pdfoutput=1

\Title{Bardeen Black Holes in the Regularized $4D$ Einstein--Gauss--Bonnet Gravity}

\TitleCitation{Bardeen Black Holes in the Regularized $4D$ Einstein--Gauss--Bonnet Gravity}

\Author{Arun Kumar $^{1,2,}$*\orcidA{}, Rahul Kumar Walia $^{1,3}$\orcidB{} and Sushant G. Ghosh $^{1,3}$\orcidC{}}

\AuthorNames{Arun Kumar, Rahul Kumar Walia and Sushant G. Ghosh}

\AuthorCitation{Kumar, A.; Walia, R.K.; Ghosh, S.G.}

\address{%
$^{1}$ \quad Centre for Theoretical Physics, Jamia Millia Islamia, New Delhi 110025, India;  sghosh2@jmi.ac.in (S.G.G.)\\
$^{2}$ \quad Department of Mathematical Science, University of Zululand, Private Bag X1001, {Kwa-Dlangezwa}~3886,~South Africa\\
$^{3}$ \quad Astrophysics Research Centre, School of Mathematics, Statistics and Computer Science, University of KwaZulu-Natal, Private Bag 54001, Durban~4000, South Africa; rahul.phy3@gmail.com (R.K.);\\}

\corres{Correspondence: arunbidhan@gmail.com}

\abstract{We obtain exact Bardeen black holes to the regularized $4D$ Einstein--Gauss--Bonnet (EGB) gravity minimally coupled with the nonlinear electrodynamics (NED). In turn, we analyze the horizon structure to determine the effect of GB parameter $\alpha$ on the minimum cutoff values of mass, $M_0$, and magnetic monopole charge, $g_0$, for the existence of a black hole horizon. We obtain an exact expression for thermodynamic quantities, namely, Hawking temperature $T_+$, entropy $S_+$, Helmholtz free energy $F_+$, and specific heat $C_+$ associated with the black hole horizon, and they show significant deviations from the $4D$ EGB case owing to NED.
Interestingly, there exists a critical value of horizon radius, $r_+^{c}$, corresponding to the local maximum of Hawking temperature, at which heat capacity diverges, confirming the second-order phase transition. A discussion on the black holes of alternate regularized $4D$ EGB gravity belonging to the scalar-tensor theory is appended.}


\begin{document}

\section{Introduction}
Lovelock's theorem \cite{Lovelock:1972vz} encapsulates that in four-dimensional ($4D$) spacetimes, the Lagrangian density leads to the second-order equation of the motion, respects the metricity, and diffeomorphism invariances are just the Ricci scalar with a cosmological constant term. Higher-order terms in the curvature invariants may lead to gravitational field equations having more than second-order derivatives of the metric, which generically at the quantum level would lead to ghosts and instabilities. However, for $D$ dimensional spacetime with $D>4$, the Lagrangian density having only a specific polynomial of higher-curvature terms can lead to the second-order equation of motion; Einstein--Gauss--Bonnet (EGB) gravity~\cite{Lanczos:1938sf} and Lovelock gravity \cite{Lovelock:1971yv} are such particular examples of a natural extension of the Einstein's general relativity in $D>4$. These theories are free from pathologies that affect other higher-derivative gravity theories and have the same degrees of freedom as Einstein's general relativity. The EGB gravity theory contains quadratic corrections in curvature tensor invariants, which also appears as the leading correction to the effective low-energy action of the heterotic string theory \cite{Zwiebach:1985uq,Duff:1986pq,Gross:1986mw}. Over the past decade, EGB gravity had a broad interest in theoretical physics and appeared to be quite compatible with the available astrophysical data \cite{Koivisto:2006xf,Amendola:2007ni}. Boulware and Deser \cite{Boulware:1985wk} obtained the first black hole solution in the $5D$  EGB gravity, and since then steady attentions have been devoted to black hole solutions, including their formation, stability, and thermodynamics \cite{Cvetic:2001bk,Padilla:2003qi,Dehghani:2006cu,Dehghani:2006ke,Dominguez:2005rt,Kobayashi:2005ch}.

This is well known that the Gauss--Bonnet (GB) correction $\mathcal{L}_{\text{GB}}$ to the Einstein--Hilbert action is a topological invariant in $D<5$, e.g., for a maximally symmetric spacetime with the curvature measure $\mathcal{K}$, the variation of the GB Lagrangian $\mathcal{L}_{\text{GB}}$ reads
\begin{linenomath}
\begin{equation}\label{gbc}
	\frac{g_{\mu\sigma}}{\sqrt{-g}} \frac{\delta \mathcal{L}_{\text{GB}}}{\delta g_{\nu\sigma}} = \frac{\alpha (D-2) (D-3)(D-4)}{2(D-1)} {\cal K}^2 \delta_{\mu}^{\nu},
\end{equation}
\end{linenomath}
where
\begin{linenomath}
\begin{equation}
\mathcal{L}_{\text{GB}}=R^{\mu\nu\rho\sigma} R_{\mu\nu\rho\sigma}- 4 R^{\mu\nu}R_{\mu\nu}+ R^2,\label{GB}
\end{equation}
\end{linenomath}
and $R$, $R_{\mu\nu}$, and ${R}{^\mu}_{\nu\gamma\delta}$ are, respectively, the Ricci scalar, Ricci tensor, and Reimann tensor, and $g$ is the determinant of the metric tensor $g_{\mu\nu}$. Further, the contribution of $\mathcal{L}_{\text{GB}}$ to all the components of Einstein's equation are also proportional to ($D-4$), therefore it does not have any contribution on the gravitational dynamics in $D=4$; where $\mathcal{K}$ is the spacetime curvature measure. However, this can be compensate by re-scaling the coupling constant by $\alpha\to \alpha/(D-4)$, and the four-dimensional theory is defined as the limit $D\to 4$ at the level of equations of motion \cite{Glavan:2019inb}. Thereby the GB action bypasses all conditions of Lovelock's theorem \cite{Lovelock:1972vz} and makes a non-trivial contribution to the gravitational dynamics even in $4D$. Since its inception, several works devoted to $4D$ EGB gravity have been reported, namely the stability, quasinormal modes, and shadows of spherically symmetric black hole~\cite{Konoplya:2020bxa,Guo:2020zmf,Ghosh:2020syx}, charged black holes in AdS space \cite{Fernandes:2020rpa,Ghosh:2020ijh}, rotating black holes and their shadows \cite{Wei:2020ght,Kumar:2020owy}, Vaidya-like radiating black holes \cite{Ghosh:2020vpc}, and relativistic stars solutions \cite{Doneva:2020ped}. The motion of a classical spinning test particle \cite{Zhang:2020qew,Kumar:2020ltt}, gravitational lensing \cite{Islam:2020xmy,Kumar:2020sag}, and thermodynamical phase transitions in AdS space \cite{Hegde:2020xlv} have also been investigated. The generalizations to the higher-order regularized
Einstein--Lovelock theory are presented in Refs.~\cite{Konoplya:2020qqh,Casalino:2020kbt}. The static spherically symmetric black hole solution in the novel $4D$ EGB gravity reads as \cite{Glavan:2019inb}
\begin{linenomath}
\begin{equation}
	ds^2=-f(r)dt^2+f(r)^{-1}dr^2+r^2 (d\theta^2+\sin^2\theta\,d\phi^2),
\end{equation} 
\end{linenomath}
with
\begin{linenomath}
\begin{equation}\label{gls}
f(r)=1+\frac{r^2}{2\alpha}\left(1\pm\sqrt{1+\frac{8M\alpha}{r^3}}\right),
\end{equation}
\end{linenomath}
such that at the short distances the gravitational force is repulsive and thus an infalling particle never reaches $r=0$, however, the curvature invariant diverges at $r=0$ \cite{Glavan:2019inb}.

The idea of $4D$ regularization of EGB gravity was pioneered by Tomozawa \cite{Tomozawa:2011gp}, and later Cognola {et al.} \cite{Cognola:2013fva} reintroduced it by accounting quantum corrections due to a GB invariant within a classical Lagrangian approach. However, several works have reported subtleties of the Glavan and Lin \cite{Glavan:2019inb} regularization procedure. Nevertheless, some consistent and standard alternate regularization procedures for $4D$ EGB gravity were also proposed, which contrary to the Glavan and Lin \cite{Glavan:2019inb} findings, mainly belong to the special class of scalar-tensor theory of gravity  \cite{Lu:2020iav,Kobayashi:2020wqy,Hennigar:2020lsl,Casalino:2020kbt,Ma:2020ufk,Arrechea:2020evj,Fernandes:2020nbq}. However, noteworthily, none of the critics~\cite{Ai:2020peo,Gurses:2020ofy,Hennigar:2020lsl} disprove dimensional regularization procedure of Glavan and Lin \cite{Glavan:2019inb} at least for the case of maximally symmetric or spherically symmetric spacetimes. It is only because this dimensional regularization procedure depends on the choice of higher-dimensional spacetime metric, and Glavan and Lin's \cite{Glavan:2019inb} procedure may not work well for spacetimes beyond the spherical symmetry that alternate regularization procedures were proposed. Moreover, it is worth mentioning that although these alternate regularization procedures are completely different in spirit, the static spherically symmetric black hole solutions of these alternate $4D$ regularized theories \cite{Ma:2020ufk,Casalino:2020kbt,Hennigar:2020lsl,Lu:2020iav} are exactly same as obtained by Glavan and Lin \cite{Glavan:2019inb} as demonstrated in Appendix \ref{Apd}.
 Furthermore, the static and spherically symmetric black hole solution~\cite{Tomozawa:2011gp,Cognola:2013fva,Glavan:2019inb} of $4D$ EGB gravity is identical as those found in semi-classical Einstein's equations with conformal anomaly \cite{Cai:2009ua,Cai:2014jea}, in gravity theory with quantum corrections~\cite{Cognola:2013fva}. In addition,   observations, namely, EHT black hole shadow \cite{Wei:2020ght,Kumar:2020owy}, cosmology, binary black hole systems, perihelion precession of Mercury, the precession of S2 around Sgr A*, gravitational waves from binary neutron star merger, and table-top experiments \cite{Clifton:2020xhc}, have been used to investigate the validity of $4D$ EGB gravity and to placed constraints on the GB coupling parameter. Moreover, studying the non-trivial higher-curvature gravity effects in $4D$ EGB black hole spacetimes is vital for better understanding the validity of the theory. 

The first ever regular black hole solution in general relativity was presented by Bardeen~\cite{Bardeen:1968} that is asymptotically flat at large $r$, and near the origin behaves as de-Sitter, such that all curvature invariants take finite values at $r=0$. Later, Ayon-Beato and Garcia~\cite{AyonBeato:1998ub} invoked non-linear electromagnetic field (NED) to show that the Bardeen black hole is an exact solution of general relativity minimally coupled to NED. Since then, significant interest developed in finding regular black hole solutions \cite{Ayon-Beato:1999qin,Dymnikova:1992ux,Dymnikova:2004zc,Bronnikov:2000vy,Bronnikov:2005gm,Burinskii:2002pz,Hayward:2005gi,Junior:2015fya,Sajadi:2017glu,Fan1:2016hvf,Bronnikov:2017tnz,Toshmatov:2018cks,Bronnikov:2000yz} and an enormous advance in the analysis and uncovering properties of regular black holes have been reported \cite{Ansoldi:2008jw,Lemos:2011dq,Bambi:2014nta,Schee:2015nua,Kumar:2020bqf,Ghosh:2020tgy,Kumar:2020cve}. Discussions of Bardeen's solution on issues of quasinormal modes, stability \cite{Fernando1:2012yw,Flachi:2012nv,Ulhoa:2013fca,Breton:2016mqh,Saleh:2018hba,Toshmatov:2019gxg,Dey:2018cws}, particle acceleration \cite{Ghosh:2015pra}, and the  thermodynamics have  been carried out \cite{Ali:2018boy,Breton:2014axa,Ali:2019myr,Ghaderi:2017yfr,Man:2013hpa,Sharif:2011ja,Sharif:2019yiy}.  Additionally, several extensions of the Bardeen black hole have been considered, which includes Bardeen de Sitter \cite{Fernando:2016ksb,Singh:2017qur}, Bardeen anti-de Sitter \cite{Li:2018bny}, Bardeen's  solution to non-commutative inspired geometry \cite{Sharif:2011ja}, and also in higher dimensional spacetimes \cite{Ali:2018boy}. Recently, Bardeen black hole solutions have also been found and discussed in the EGB gravity theory \cite{Kumar:2018vsm,Singh:2019wpu}.  

This paper presents the spherically symmetric static Bardeen-like regular black holes of the regularized $4D$ EGB gravity coupled with the NED field. We also showed an identical solutions in the physically motivated alternate regularized $4D$ EGB gravity, based on the Kaluza--Klein-like dimensional reduction procedure. Three parameters characterize the black hole metric, mass $M$, GB coupling parameter $\alpha$, and the magnetic charge $g$ coming from the NED. We discuss the structure of black hole horizons, which depending on the values of black hole parameters, can describe the extremal black holes with degenerate horizons and nonextremal black holes with two distinct horizons. Analytic expressions for the various thermodynamical quantities, including the temperature, free energy, and specific heat, are obtained. The paper is organized as follows: We begin in Section~\ref{BAdS_section} with the construction of the spherically symmetric black holes in novel $4D$ EGB gravity coupled with the NED. Section~\ref{sec3} is devoted to the study of black hole thermodynamics and phase transitions. Finally, we summarize our main findings in  Section~\ref{sec4}.

\section{Bardeen Black Holes in $4D$ Einstein--Gauss--Bonnet Gravity: Exact Solutions and Horizons}
\label{BAdS_section}

The action for the EGB gravity minimally coupled with the NED field in the  $D$-dimensional spacetime, reads \cite{Cai:2001dz,Ghosh:2020tgy,Kumar:2018vsm}
\begin{linenomath}
\begin{equation}\label{action}
\mathcal{I}_{G}=\int d^{D}x\sqrt{-g}\left[\frac{1}{16\pi}\Big (R +\alpha' \mathcal{L}_{\text{GB}}\Big)-\frac{1}{4\pi}\mathcal{L(F)} \right],
\end{equation}
\end{linenomath}
where $\alpha'=\alpha/(D-4)$ is re-scaled GB coupling constant, and the Lagrangian density $\mathcal{L(F)}$ is a function of invariant $\mathcal{F}=F^{\mu\nu}F_{\mu\nu}/4$ with $F_{\mu\nu}=\partial_{\mu}A_{\nu}-\partial_{\nu}A_{\mu}$ being the electromagnetic field tensor for the gauge potential $A_{\mu}$. Varying the action (\ref{action}) with metric tensor $g_{\mu\nu}$, yields the equations of gravitational field as follows
\begin{linenomath}
\begin{equation}
G_{\mu\nu}+\alpha' H_{\mu\nu}=T_{\mu\nu}\equiv 2\Bigr(\mathcal{L_F} \mathcal{F}_{\mu\sigma}\mathcal{F}_{\nu}{^\sigma}-g_{\mu\nu}\mathcal{L(F)} \Bigl),
\end{equation}
\end{linenomath}
where 
\begin{linenomath}
\begin{eqnarray}
G_{\mu\nu}&=&R_{\mu\nu}-\frac{1}{2}R g_{\mu\nu},\nonumber\\
H_{\mu\nu}&=&2\Bigr( R R_{\mu\nu}-2R_{\mu\sigma} {R}{^\sigma}_{\nu} -2 R_{\mu\sigma\nu\rho}{R}^{\sigma\rho} - R_{\mu\sigma\rho\delta}{R}^{\sigma\rho\delta}{_\nu}\Bigl)-\frac{1}{2}\mathcal{L}_{\text{GB}}g_{\mu\nu},\label{FieldEq}
\end{eqnarray}
\end{linenomath}
and $T_{\mu\nu}$ is the energy momentum tensor for the NED field. We consider the Lagrangian density for the NED field in the $D$ dimensional spacetime \cite{Ali:2018boy,Kumar:2018vsm}
\begin{linenomath}
\begin{equation}
\mathcal{L(F)}=\frac{(D-1)(D-2)\mu'^{D-3}}{4g^{D-1}}\left(\frac{\sqrt{2g^2 \mathcal{F}}}{1+\sqrt{2g^2 \mathcal{F}}} \right)^{\frac{2D-3}{D-2}}, \label{ned}
\end{equation}
\end{linenomath}
where $$\mathcal{F}=\frac{g^{2(D-3)}}{2r^{2(D-2)}}.$$
Here, $g$ is the magnetic monopole charge of NED. We are using Lagrangian density associated with magnetic monopole charge because this kind of NED field generates a globally regular black hole solutions. Whereas the Lagrangian density associated with electric charge changes with the change in radial coordinate, the Lagrangian density associated with electric charge can not produce regular black holes globally \cite{Bronnikov:2000vy}. Additionally, regular black holes sourced by the electric charge usually do not recover the Maxwell linear electrodynamics in the weak field limits. The $D$-dimensional static, spherically symmetric metric ansatz~\cite{Ghosh:2020tgy,Kumar:2018vsm}~reads 
\begin{linenomath}
\begin{equation}
ds^2=-f(r)dt^2+f(r)^{-1}dr^2+r^2d\Omega_{D-2}^2,\label{metric}
\end{equation} 
\end{linenomath}
with $f(r)$ is the metric function to be determined and
\begin{linenomath}
\begin{equation}
d\Omega^2_{D-2}=d\theta_1^2+\sum_{i=2}^{D-2}\left[\prod_{j=2}^i \sin^2\theta_{j-1} \right]d\theta^2_i,
\end{equation} 
\end{linenomath}
is the line element of a $(D-2)$-dimensional unit sphere \cite{Myers:1986un,Xu:1988ju}. Using field equations (\ref{FieldEq}) and metric ansatz (\ref{metric}), in the limit $D\to 4$ we get the following $(r,r)$ equation  \cite{Ghosh:2014pga}
\begin{linenomath}
\begin{eqnarray}
r^3f^{\prime}(r)+ \alpha\big(f(r)-1\big)\Big(f(r)-1-2rf^{\prime}(r)\Big) + r^2\big(f(r)-1\big)= -\frac{6\mu'g^2 r^4}{(r^2+g^2)^{5/2}},\label{rr1}
\end{eqnarray} 
\end{linenomath}
which yields the following solution
\begin{linenomath}
\begin{equation}
f_{\pm}(r)=1+\frac{r^2}{2\alpha}\left(1\pm\sqrt{1+\frac{8M\alpha}{(r^2+g^2)^{3/2}}}\right),\label{fr}
\end{equation}
\end{linenomath}
where $\mu'=M$ is identified as the black hole mass parameter. 
The metric Equation~(\ref{metric}) for $D\to 4$ and  $f(r)$ in Equation~(\ref{fr}) describes a static spherically symmetric Bardeen-like regular black hole in $4D$ EGB gravity coupled with the NED field, which in the limits of $g=0$ goes over to the black hole solution of Ref.~\cite{Glavan:2019inb}. However, for the special case of vanishing GB coupling, $\alpha\to 0$, the metric (\ref{metric}) retains the spherically symmetric Bardeen black hole \cite{Bardeen:1968}.
The ``$\pm$'' sign in Equation~(\ref{fr}) refers to two different branches of solutions. To analyze the general structure of the solution, we take the asymptotically large $r$ limit, $r\to \infty$, to obtain
\begin{linenomath}
\begin{eqnarray}
&&\lim_{r\to \infty}f_+(r)=1+\frac{2M}{r} +\frac{r^2}{\alpha}+\mathcal{O}\Big(\frac{1}{r^3}\Big),\nonumber\\
&&\lim_{r\to \infty}f_-(r)=1-\frac{2M}{r} +\mathcal{O}\Big(\frac{1}{r^3}\Big).
\end{eqnarray} 
\end{linenomath}

One can easily see that the negative branch of the solution goes over to the well-known Schwarzschild black hole solution, whereas the positive branch does not correspond to any physical case when $\alpha\to 0$. Boulware and Deser \cite{Boulware:1985wk} have also shown that $5D$ EGB black holes with the ``$+$'' branch sign are unstable, and the graviton degree of freedom is a ghost, while the branch with the ``-'' sign is stable and is free of ghosts. Hence, we will take only the negative branch $f_-(r)$ of the solution hereafter. Furthermore, it is legitimate to give $4D$ EGB solutions a physical interpretation, investigate their properties, and compare them with the corresponding ones in Einstein's gravity can then help shed some light on the validity of $4D$ EGB gravity.

The presence of coordinate singularity of the metric Equation~(\ref{metric}) at $f(r)=0$ signifies the existence of the horizons, which are the real and positive roots of the equation
\begin{linenomath}
\begin{equation}
(r^2+g^2)^{3/2}(r^2+\alpha)-2Mr^4=0,\label{horizon}
\end{equation}
\end{linenomath}

\begin{figure}[H]
\includegraphics[width=7 cm]{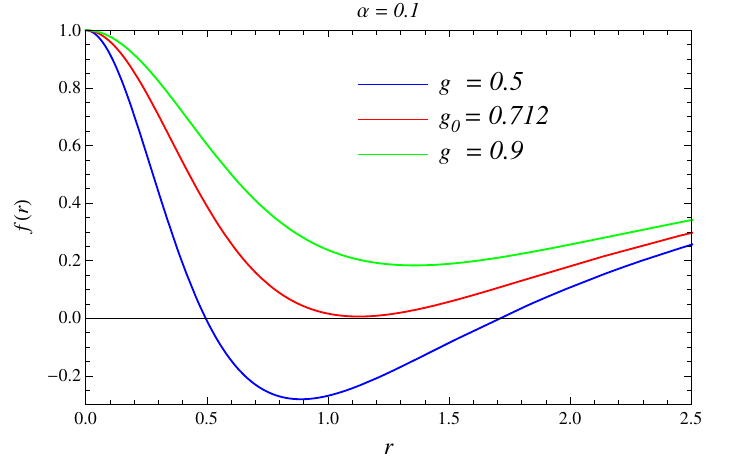}
\includegraphics[width=7 cm]{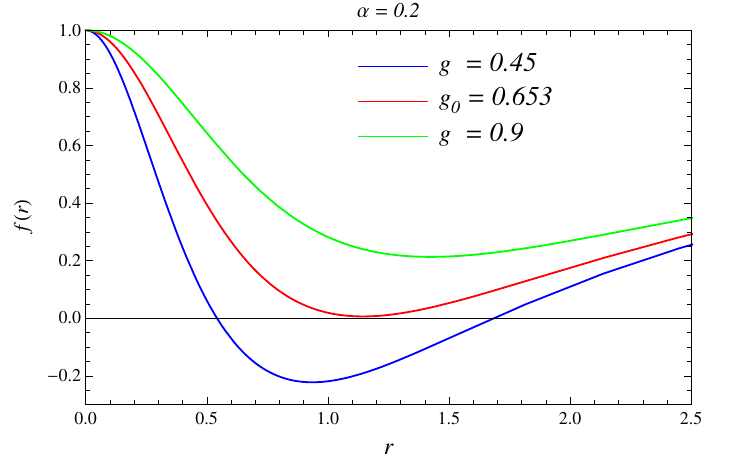}\\
\includegraphics[width=7 cm]{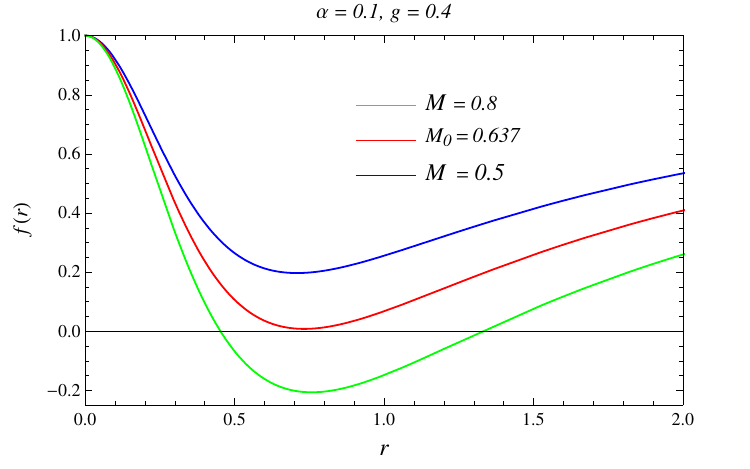}
\includegraphics[width=7 cm]{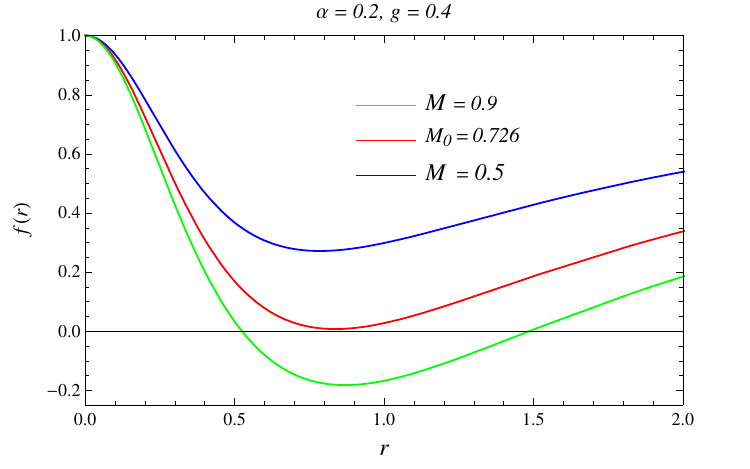}\\
\includegraphics[width=7 cm]{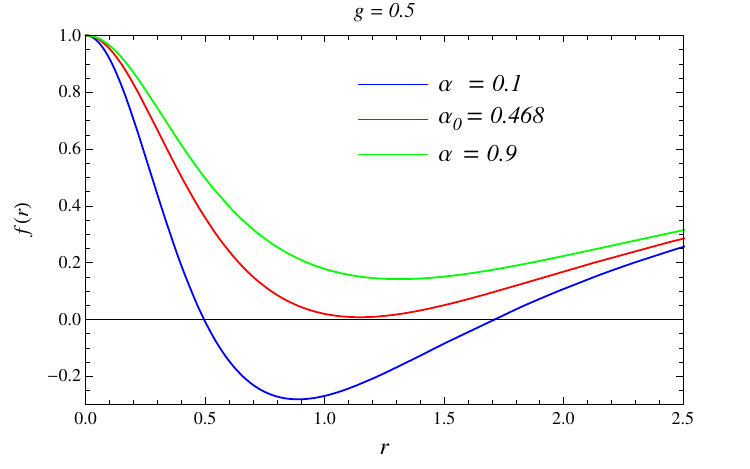}
\includegraphics[width=7 cm]{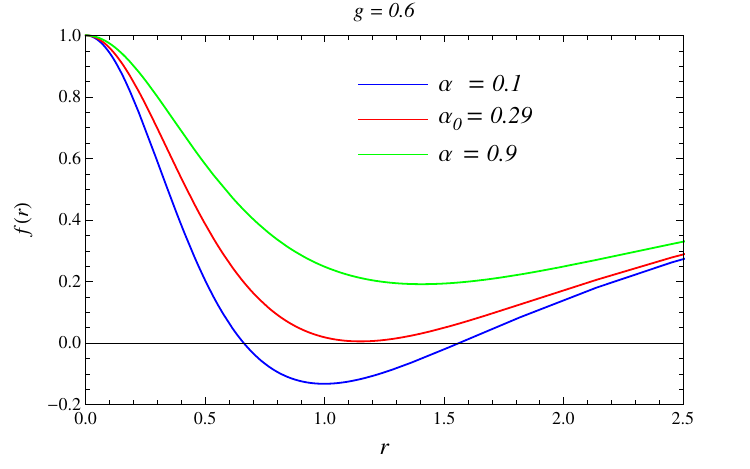}
\caption{Metric function $f(r)$ \textit{vs} $r$ for Bardeen black holes for various values of parameters.}
\label{fig:f1}
\end{figure}

 From Figure~\ref{fig:f1}, one can notice that for given values of $g$ and $\alpha$ (or $M$ and $\alpha$), there exists an extremal value $M_0$ (or $g_0$), such that for $M>M_0$ (or $g<g_0$), there exist two distinct horizons $r_{\pm}$ and for  $M<M_0$ (or $g>g_0$), there are no horizons. Similarly, for given values of $M$ and $g$, one can find the extremal value of $\alpha=\alpha_0$. Here, $r_-$ and $r_+$, respectively, represent the Cauchy and the event horizon. For extremal value of parameters, namely $M=M_0$ or $g=g_0$ or $\alpha=\alpha_0$, the two horizons $r_{\pm}$, coincide and leads to an extremal black hole with degenerate horizon $r_+=r_-\equiv r_E$. It can be noticed that the maximum (or minimum) cutoff value of charge $g_0$ (or $M_0$) increases (decreases) as we increase the value of GB coupling constant $\alpha$ (cf. Figure~\ref{fig:f1}).

\section{ Black Hole Thermodynamics}
\label{sec3}
Next, we obtain exact expressions for the thermodynamical quantities like, Arnowitt--Deser--Misner mass ($M_+$), Hawking temperature ($T_+$), entropy ($S_+$), Helmholtz's free energy ($F_+$), and heat capacity ($C_+$) associated with the black hole horizon $r_+$. The solution of equation $f(r_+)=0$, where $r_+$ is the event horizon radius, leads  to the mass 
\begin{linenomath}
\begin{equation} \label{B_mass}
M_+= \frac{r_+}{2} \left[\Big(1+\frac{\alpha}{r_+^2}\Big)\Big(1+\frac{g^2}{r^2}\Big)^{\frac{3}{2}}\right],
\end{equation}
\end{linenomath} which in the absence of charge $g$, goes to the value of mass for $4D$ EGB black holes \cite{Fernandes:2020rpa}
\begin{linenomath}
\begin{equation} \label{B_mass1}
M_+= \frac{r_+}{2} \left[1+\frac{\alpha}{r_+^2}\right],
\end{equation}
\end{linenomath}whereas in the limit of vanishing coupling parameter $\alpha=0$, reduces to mass of Schwarzschild black holes \cite{Cho:2002hq,Kumar:2018vsm} 
\begin{linenomath}
\begin{equation} \label{B_mass1}
M_+= \frac{r_+}{2}.
\end{equation}
\end{linenomath}

The surface gravity, $\kappa$=$\sqrt{-\frac{1}{2}\bigtriangledown_{\mu}\chi_{\nu}\bigtriangledown^{\mu}\chi^{\nu}}$, at the horizon of black hole reads as
\begin{linenomath}
\begin{equation}
\kappa=\frac{1}{2\pi}\frac{\partial f(r)}{\partial r}|_{r=r_+}=\frac{1}{2r_+}\left[\frac{r_+^2(r_+^2-\alpha)-2g^2(r_+^2+2\alpha)}{(r_+^2+g^2)(r_+^2+2\alpha)}\right].
\end{equation} 
\end{linenomath}

Now, by using the relation $T_+=\kappa/2\pi$ \cite{Bardeen:1973gs,Bekenstein:1973ur}, we get the expression for the Hawking temperature associated with black hole horizon, which reads
\begin{linenomath}
\begin{equation} \label{B_temp0}
T_+= \frac{1}{4\pi r_+}\left[\frac{r_+^2(r_+^2-\alpha)-2g^2(r_+^2+2\alpha)}{(r_+^2+g^2)(r_+^2+2\alpha)}\right].
\end{equation}
\end{linenomath}

For the $g=0$, Equation~(\ref{B_temp0}) reduces to the temperature of $4D$ EGB black holes \cite{Fernandes:2020rpa} 
\begin{linenomath}
\begin{equation} \label{B_temp1}
T_+= \frac{1}{4\pi r_+}\left[\frac{r_+^2-\alpha}{r_+^2+2\alpha}\right],
\end{equation}
\end{linenomath}
and further in the limit of $\alpha\to 0$, we obtained the temperature of Schwarzschild black holes \cite{Cho:2002hq,Kumar:2018vsm}
\begin{linenomath}
\begin{equation}
T_+= \frac{1}{4\pi r_+}.
\end{equation}
\end{linenomath}

\begin{figure}[H] 
\includegraphics[width=7 cm]{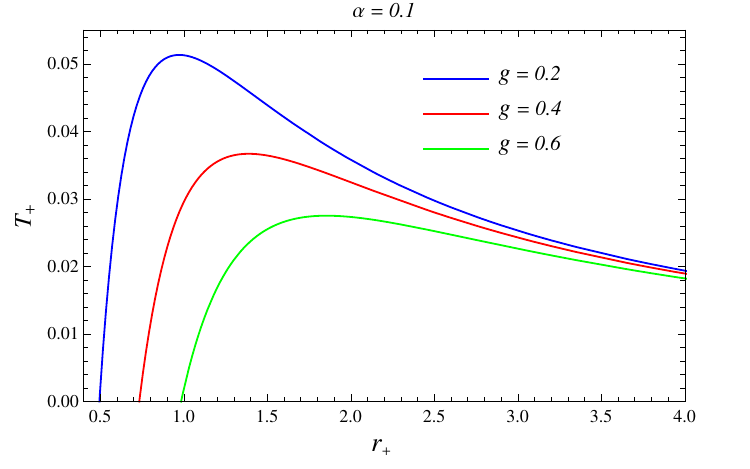}
\includegraphics[width=7 cm]{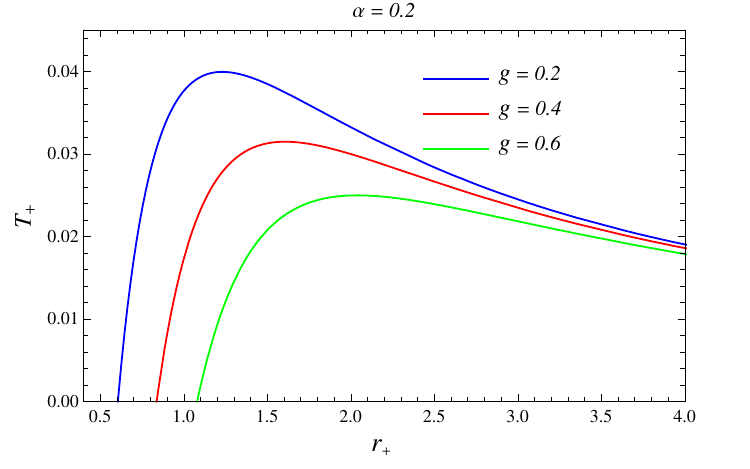}\\
\includegraphics[width=7 cm]{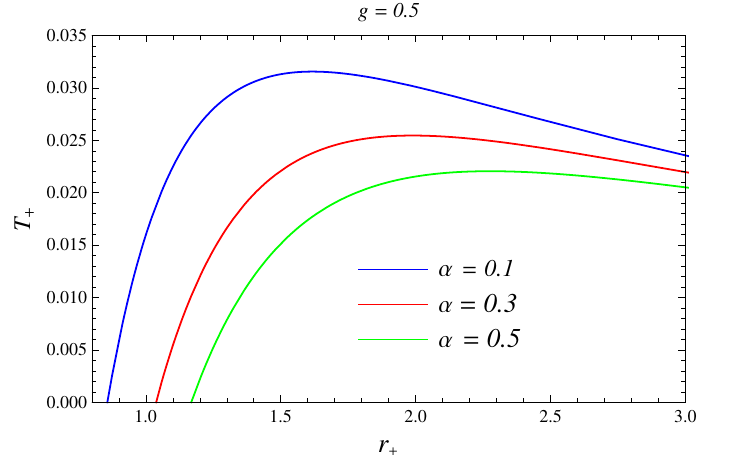}
\includegraphics[width=7 cm]{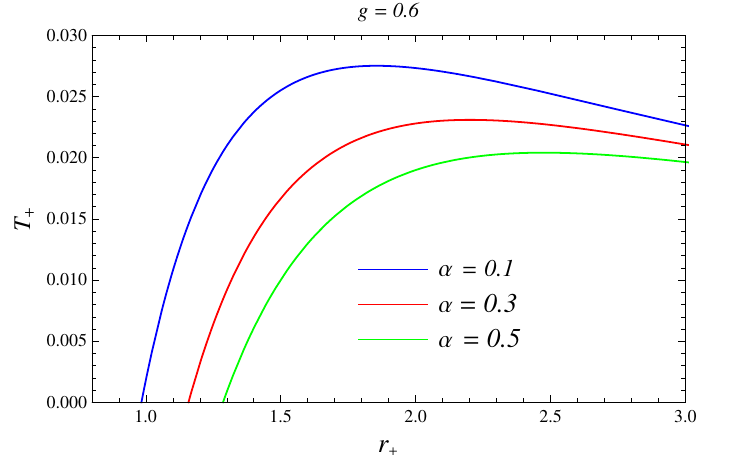}
\caption{ Hawking temperature ${T_+}$ \textit{vs} horizon radius $r_+$ for various values of $g$ and $\alpha$.}
\label{fig:Btemp}
\end{figure}

We depict the Hawking temperature behavior with horizon radius  $r_+$ for different values of $g$ and $\alpha$ in Figure~\ref{fig:Btemp}. It is evident that, during the Hawking evaporation, as the black hole horizon size shrinks, the temperature initially increases with decreasing $r_+$, reaches a maximum value, and then rapidly decreases and eventually vanishes for some particular value of $r_+$. The vanishing temperature implies the extremal black hole with degenerate horizons, such that $T_+<0$ corresponds to the no-black hole states. The Hawking temperature possesses a local maximum at a particular value of horizon radius, say $r_+^c$ -- critical horizon radius. Since the Hawking temperature has the maximum value at $r_+^c$, the first derivative of the temperature vanishes at a critical radius, leading to the divergence of specific heat. 
The value of critical horizon radius increases with magnetic monopole charge $g$ and GB coupling constant $\alpha$ (cf.
 Figure~\ref{fig:Btemp}).
By taking into account the first law of black hole thermodynamics, $dM_+=T_+dS_+$, we obtain the entropy of the black hole as \cite{Bardeen:1973gs,Ghosh:2008jca}
\begin{adjustwidth}{-\extralength}{0cm}
\begin{eqnarray} \label{B_s}
S_+&=& \frac{A}{4} \left[\sqrt{1+\frac{g^2}{r_+^2}}\left(1-\frac{16}{3}\frac{\alpha}{r_+^2}+\frac{2g^2}{3r_+^2}(3r_+^2+2\alpha)\right) +\frac{1}{r_+^2}(3g^2+4\alpha)\log(r_++\sqrt{g^2+r_+^2})\right],
\end{eqnarray}
\end{adjustwidth}
where $A=4\pi r_+^2 $ and for the Bardeen black holes in the $4D$ EGB gravity gets modified due to the presence of GB coupling constant $\alpha$ and charge $g$. It is worthwhile to mention that black hole entropy Equation~(\ref{B_s}) does not follows the usual entropy-area law, $S_+=A/4$, where $A=4\pi r_+^2$ is the black hole event horizon area. For $g=0$, we get
\begin{linenomath}
\begin{eqnarray} \label{B_s1}
S_+&=&\frac{A}{4}\left[1-\frac{16}{3}\frac{\alpha}{r_+^2}+\frac{4\alpha}{r_+^2}\log(2r_+)\right],
\end{eqnarray}
\end{linenomath}
and further in the limit of $\alpha=0$, it reduces to 
\begin{linenomath}
\begin{equation}
S_{+}=\frac{A}{4}.
\end{equation} 
\end{linenomath}

It is precisely the entropy of the Schwarzschild black hole and the area law holds. It is imperative to find the regions of the global stability of a thermodynamical system. Hence, we further study the behavior of Helmholtz's free energy ($F_+$) of our model. Helmholtz's free energy determines the global stability of a thermodynamical system, such that a thermodynamical system with $F_+<0$ is globally stable, whereas the system with $F_+>0$ is globally unstable. The Helmholtz's free energy of Bardeen black holes in $4D$ EGB gravity can be obtained by using the relation, $F_+=M_+-T_+S_+$, \cite{Kumar:2018vsm} as
\vspace{-6pt}

\begin{adjustwidth}{-\extralength}{0cm}
\begin{linenomath}
\begin{eqnarray}
F_+&=&\frac{r_+}{4}\Bigg[2(1+\frac{\alpha}{r_+^2})(1+\frac{g^2}{r^2})^{\frac{3}{2}}-\frac{r_+^2(r_+^2-\alpha)-2g^2(r_+^2+2\alpha)}{3r_+^4\sqrt{r_+^2+g^2}(r_+^2+2\alpha)}\nonumber\\
&&\times\left(3r_+^2-16\alpha+2g^2(3r_+^2+2\alpha)+\frac{3r_+^2}{\sqrt{r_+^2+g^2}}(3g^2+4\alpha)\log(r_++\sqrt{r_+^2+g^2})\right)\Bigg], \nonumber\\
\end{eqnarray}
\end{linenomath}
\end{adjustwidth}
for $g=0$, we get the free energy of $4D$ EGB black hole
\begin{linenomath}
\begin{eqnarray}
F_+=\frac{r_+}{4(r_+^2+2\alpha)}\left[2(1+\frac{\alpha}{r_+^2})(r_+^2+2\alpha)-(r_+^2-\alpha)\left(1-\frac{16\alpha}{3r_+^2}+\frac{4\alpha}{r_+^2}\log(2r_+)\right)\right],
\end{eqnarray} 
\end{linenomath}
which, in the limit of $\alpha\to 0$, goes over to the Helmholtz's free energy of Schwarzschild black holes \cite{Kumar:2018vsm},
\begin{linenomath}
\begin{eqnarray}
F_{+}&=&\frac{r_+}{4}.
\end{eqnarray}
\end{linenomath}

\begin{figure}[H]
\includegraphics[width=7 cm]{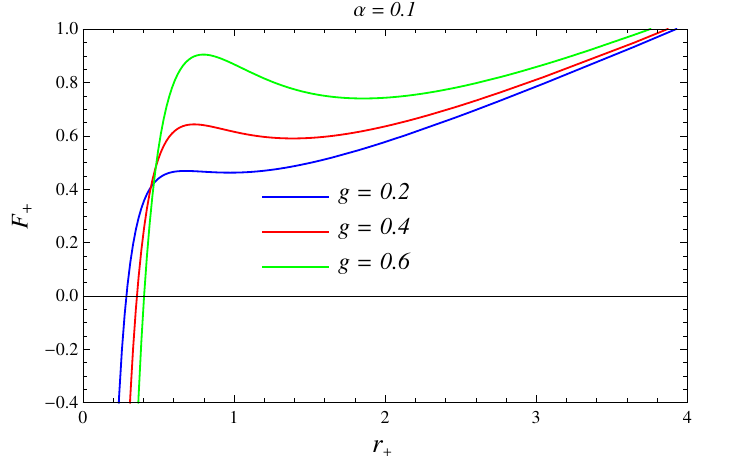}
\includegraphics[width=7 cm]{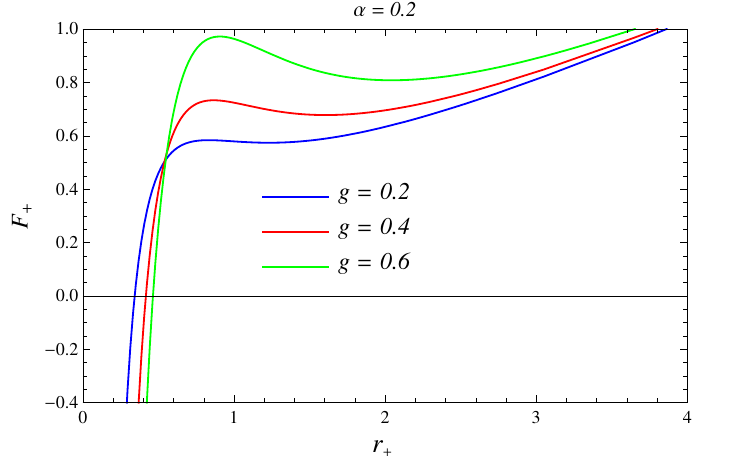}\\
\includegraphics[width=7 cm]{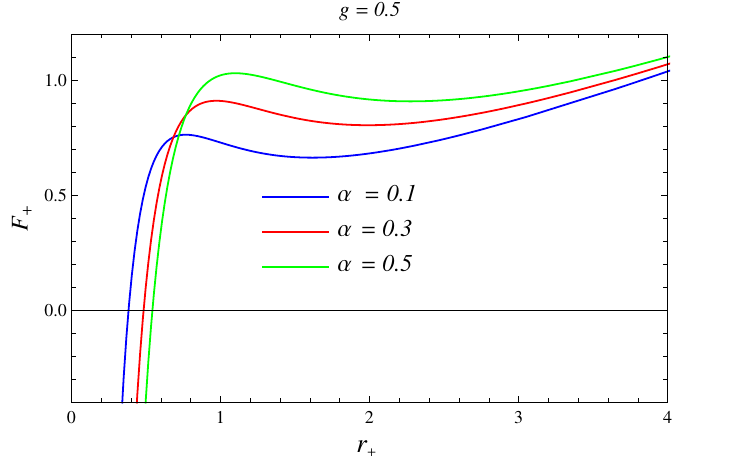}
\includegraphics[width=7 cm]{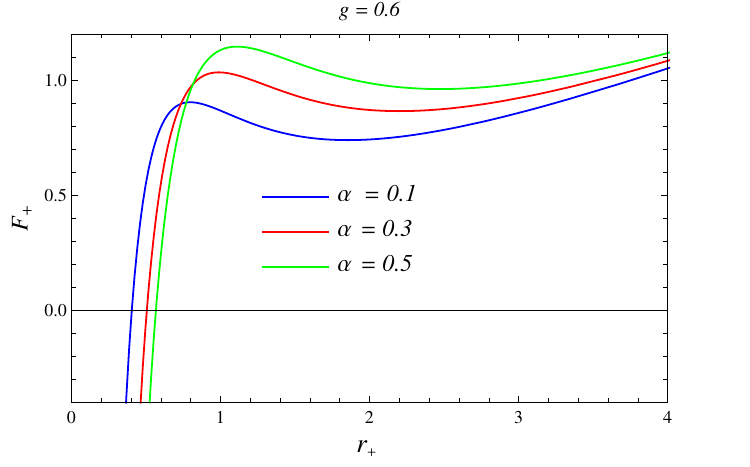}
\caption{ The Helmholtz free energy ${F_+}$ \textit{vs} horizon $r_+$ for different values of $g$ and $\alpha$.}
\label{fig:Bf}
\end{figure}

The behavior of Helmholtz's free energy is shown in Figure~\ref{fig:Bf}. However, Helmholtz's free energy is negative ($F_+<0$) for some small values of $r_+$, and the black hole temperature is negative for these states. However, for black hole states with $T_+\geq 0$, Helmholtz's free energy is always positive ($F_+>0$). Therefore, Bardeen $4D$ EGB black holes are globally unstable as $F_+>0$.  
Further, to check the local thermodynamical stability of black holes, we analyze the behavior of specific heat ($C_+$) of the black holes. The positive (negative) specific heat signifies the black holes' local thermodynamical stability (instability). By using the relation, $C_+=\frac{\partial M_+}{\partial r_+}/\frac{\partial T_+}{\partial r_+}$ \cite{Cai:2001dz,Ghosh:2014pga}, we get the expression of specific of Bardeen black holes in $4D$ EGB gravity  
\begin{linenomath}
\begin{equation} \label{BAdSCp}
C_+ =-2\pi r_+^2\left[\frac{(1+\frac{g^2}{r_+^2})(r_+^2+2\alpha)^2\left(1-\frac{\alpha}{r_+^2}-2\frac{g^2}{r_+^2}(r_+^2+2\alpha)\right)}{r_+^4-\alpha(5r_+^2+2\alpha)-A\frac{g^2}{r_+^2}-B\frac{g^4}{r_+^4}}\right],
\end{equation}
\end{linenomath}with
\begin{linenomath}
\begin{equation}
A=7r_+^4+\alpha(31r_+^2+22\alpha)~~~~~~~\text{and}~~~~~~~B=2(r_+^2+2\alpha)^2.\nonumber
\end{equation}
\end{linenomath}

In the absence of magnetic charge, $g=0$, we obtain the heat capacity of $4D$ EGB black~holes
\begin{linenomath}
\begin{equation}
C_+ =-2\pi r_+^2\left[\frac{(r_+^2+2\alpha)^2\left(1-\frac{\alpha}{r_+^2}\right)}{r_+^4-\alpha(5r_+^2+2\alpha)}\right],
\end{equation}
\end{linenomath}
which in the limit of $\alpha=0$ retain the following value
\begin{linenomath}
\begin{equation} \label{BAdSCp1}
C_+ = -2 \pi  r_+^2,
\end{equation}
\end{linenomath}
which can be identified as the value for the Schwarzschild black hole  \cite{Cho:2002hq,Kumar:2018vsm}.

\begin{figure}[H]
\includegraphics[width= 7 cm]{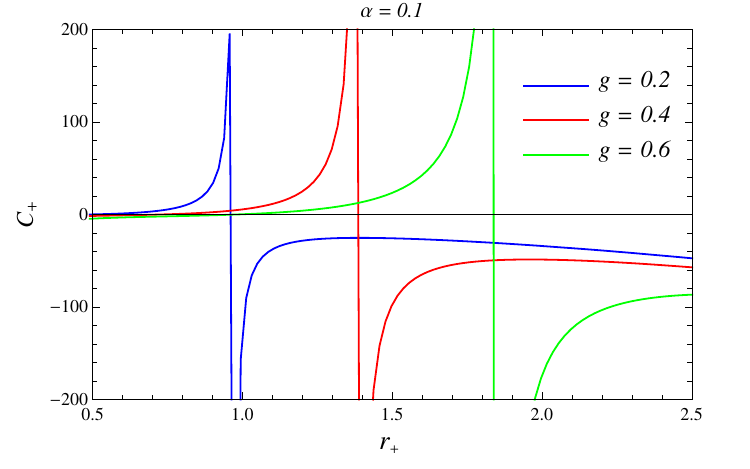}
\includegraphics[width=7 cm]{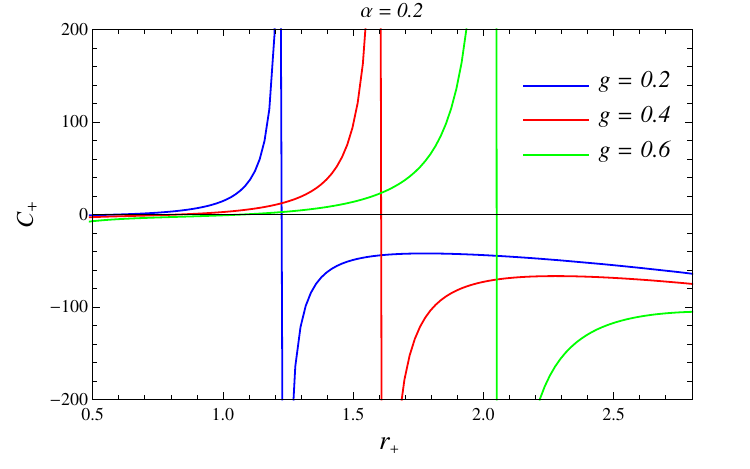}\\
\includegraphics[width=7 cm]{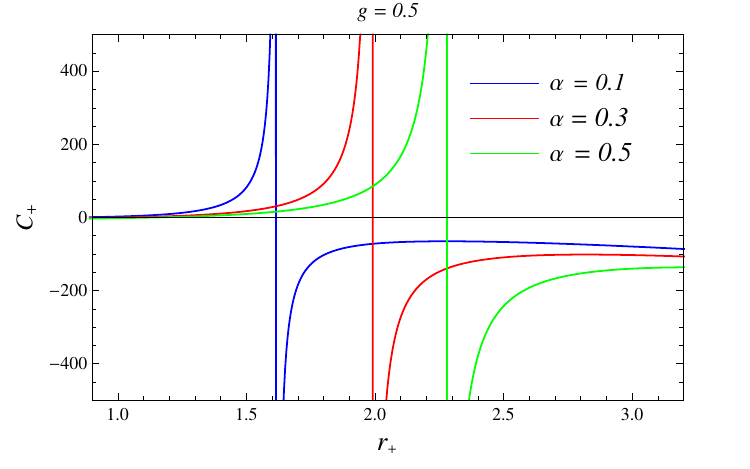}
\includegraphics[width=7 cm]{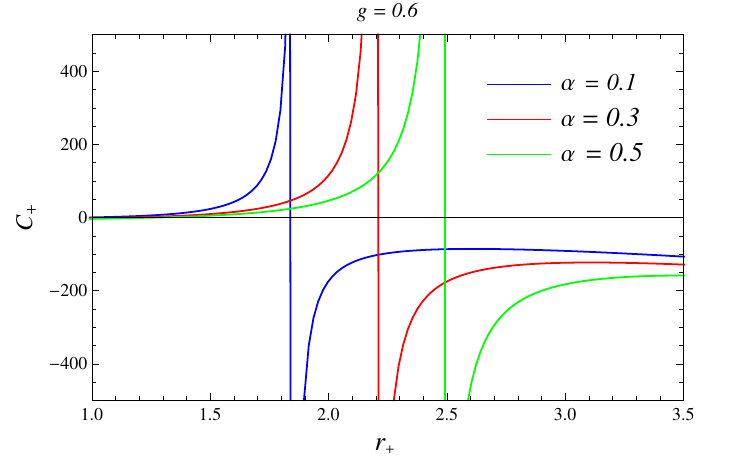}
\caption{Specific heat ${C_+}$ \textit{vs} horizon $r_+$ for different values of $g$ and $\alpha$.}
\label{fig:c}
\end{figure}

The numerical results of specific heat ($C_+$) for varying horizon radius for different values of $g$ and $\alpha$ of Bardeen black holes in $4D$ EGB gravity is shown in Figure~\ref{fig:c}. Specific heat diverges and flips its sign from positive to negative at a critical radius $r_+^c$; hence the black hole exhibits a second-order phase transition which takes the black hole from a local stable state to an unstable state. Hence, the black holes with smaller horizon radius $r_+<r_+^c$ are locally stable, whereas the black holes with larger horizon radius $r_+>r_+^c$ are locally unstable. Further, a black hole is always globally thermodynamically unstable for all $r_+$. Moreover, the black hole temperature increases with increasing $r_+$ for $r_+< r_+^c$, whereas it decreases for $r_+> r_+^c$ (cf. Figure~\ref{fig:Btemp}). Interestingly, the value of the critical radius $r_+^c$ increases as we increase the value of $g$ and $\alpha$.

\section{Discussion}\label{sec4}
In the absence of a well-defined theory of quantum gravity,  attention is devoted to phenomenological approaches to solve singularity problems arising in classical general relativity/EGB theory and studying possible implications. So an essential line of research to understand the inside of a black hole is tantamount to investigating classical black holes and their consequences, with regular, i.e., nonsingular, properties.
In view of this, we have examined static spherically symmetric Bardeen black hole solution in the EGB gravity theory in $4$-dimensional spacetime by re-scaling the GB coupling constant $\alpha \to \alpha/(D-4)$ at the level of the equation of motion with source as the NED field, and it belongs to the class of the three-parameter family of static and spherically symmetric black holes. We analyzed the black hole horizon structure and found that for a fixed value of GB coupling constant $\alpha$ and magnetic monopole charge $g$, there is a minimal value of mass, $M_0$, below which no black hole solution exists. For $M=M_0$, black holes possess degenerate horizons, and two distinct horizons for $M>M_0$. Similarly, we find the extremal values of $g=g_0$ and $\alpha=\alpha_0$ for which black hole admits degenerate horizons $r_-=r_+=r_E$.
 
We made a detailed analysis of Helmholtz free energy $F_+$ and heat capacity $C_+$ to check the thermodynamical stability of the black holes. We found that Bardeen black holes in $4D$ EGB gravity are globally thermodynamically unstable as Helmholtz's free energy is always positive. While analyzing the local thermodynamical stability through specific heat $C_+$, we found that there exists a critical value of horizon radius $r_+^{c}$ at which black holes show second-order phase transition with diverging specific heat. The black holes with $r_+<r_+^{c}$ are locally stable with $C_+>0$, whereas the black holes with $r_+>r_+^c$ having negative specific heat are locally unstable. 

Further, considering the AdS background for these $4D$ EGB Bardeen black holes may lead to exciting phenomena, like, phase structure, particle production rates, greybody factor, and critical phenomena. The possibility of further generalizing these results to $4D$ Lovelock gravity is an interesting problem for the future. Such investigations now have a clear astrophysical relevance; they can be fascinating from a future~perspective.

\textls[-15]{Furthermore,  it could be considered an effective
theory with a specific~prescription.}

\vspace{6pt} 



\authorcontributions{Conceptualization,  S.G.G.; methodology, A.K and R.K.W. ; software, A.K. and R.K.W.; validation, A.K., R.K.W. and S.G.G.; formal analysis, S.G.G., A.K and R.K.W.; investigation, S.G.G and A.K.; writing---original draft preparation, S.G.G, A.K. and R.K.W.; writing---review and editing, S. G. G., A.K. and R.K.W.; visualization, A.K. and R.K.W. ; supervision, S.G.G.; project administration, S.G.G. All authors have read and agreed to the published version of the manuscript.}


\funding{This research received no external funding.}

\institutionalreview{Not applicable.}

\informedconsent{Not applicable.}

\dataavailability{Not applicable.} 

\conflictsofinterest{The authors declare no conflict of interest.} 

\appendixtitles{yes} 

\appendix
\section[\appendixname~\thesection]{Black Holes of Alternate Regularized $4D$ EGB Gravity}\label{Apd}

Glavan and Lin \cite{Glavan:2019inb} obtained $4D$ EGB black hole solution (\ref{gls}). Here, we derive the solution by alternate regularization techniques and check if we get the same solution by them. We use the Kaluza--Klein-like dimensional compactification of the $D$-dimensional EGB gravity on a $(D-4)$-dimensional maximally symmetric space \cite{Lu:2020iav,Kobayashi:2020wqy} for obtaining the well-defined action principle for the regularized $4D$ EGB gravity minimally coupled with NED. This model is based on the idea of Mann and Ross \cite{Mann:1992ar} for obtaining the $D\to2$ limit of Einstein GR and leads to the scalar-tensor theory of gravity that belongs to a family of Horndeski gravity. Following \cite{Lu:2020iav}, we reduce the $D$-dimensional EGB gravity~(\ref{action}) to $(D-p)$ dimensional maximally symmetric space of curvature proportional to $\lambda$:
\begin{equation}
ds_D^2=ds_p^2+\exp[2\Phi]d\Sigma^2_{D-p},
\end{equation}
where $ds_p^2$ is the $p$-dimensional line element, $d\Sigma^2_{D-p}$ is the line element on the internal maximally symmetric space and $\Phi$ is the scalar field. We rescaled the GB coupling as $\alpha\to \alpha/(D-p)$ and  $4$-dimensional reduced EGB gravitational action reads as
\vspace{-10pt}

\begin{adjustwidth}{-\extralength}{0cm}
\begin{linenomath}
\begin{eqnarray}
\mathcal{I}_4=&\int d^4x\sqrt{-g}\Big[\mathcal{R}
+\alpha\Big(\Phi\,\mathcal{L}_{GB}+4G^{\mu\nu}\partial_\mu\Phi\partial_\nu\Phi-2\lambda \mathcal{R}e^{-2\Phi} -4(\partial\Phi)^2\Box \Phi+2\left((\partial\Phi)^2\right)^2\nonumber\\
&-12\lambda(\partial\Phi)^2e^{-2\Phi}-6\lambda^2e^{-4\Phi}\Big)-4\mathcal{L}(\mathcal{F})\Big],\label{action2}
\end{eqnarray}
\end{linenomath}
\end{adjustwidth}
and corresponds to the regularized $4D$ EGB gravity action with rescaled GB coupling constant. One can obtain the covariant field equations by varying the action (\ref{action2}) for metric tensor $g_{\mu\nu}$ and scalar field $\Phi(r)$ \cite{Hennigar:2020lsl}. To study the static spherically symmetric black hole solution, we consider the metric \textit{ansatz} and scalar field as follows
\begin{linenomath}
\begin{equation}
ds_4^2=-\exp[-2\chi(r)]f(r)dt^2+\frac{dr^2}{f(r)}+r^2d\Omega^2_2,\quad \Phi=\Phi(r).\label{anstaz}
\end{equation}
\end{linenomath}

On substituting the ansatz (\ref{anstaz}) to action $\mathcal{I}_4$ in (\ref{action2}), we obtain the effective Lagrangian  

\begin{adjustwidth}{-\extralength}{0cm}
\begin{linenomath}
\begin{eqnarray}
L_{\rm eff}&=&e^{-\chi}\Big[2(1- f- r f') +\frac{2}{3}\Big(
3 r^2 f^2 \Phi '^3+2 r  \left(-r f'+2 r f \chi '-4 f\right)f \Phi '^2\nonumber\\
&&-6  \left(-r f'+2 r f \chi '-f+1\right)f \Phi '-6 (f-1) \left(f'-2 f \chi '\right)\Big)\alpha\Phi'\nonumber\\
&&+ 4\alpha \lambda e^{-2\Phi} \Big(r^2 f' \Phi '-2 r^2 f \chi ' \Phi '-3 r^2 f \Phi '^2+r f'+f-1\Big)-6\alpha\lambda^2 r^2 e^{-4\Phi}-4r^2\mathcal{L}(\mathcal{F})\Big].
\end{eqnarray}
\end{linenomath}
\end{adjustwidth}

On using the Euler--Lagrange equations, we obtain the dynamical equations for metric functions $f(r)$ and $\chi(r)$, and scalar field $\Phi(r)$. Considering the special case of $\chi(r)=0$ \cite{Lu:2020iav}, these equations for the internally flat spacetime ($\lambda=0$), respectively, read as

\begin{adjustwidth}{-\extralength}{0cm}
\begin{linenomath}
\begin{eqnarray}
&&\exp[\Phi] \alpha \Big(1 -(1 - r \Phi')^2f\Big) (\Phi'^2+\Phi'')=0,\label{A1}\\
&&\exp[3\Phi]\alpha\Bigg[ \Big(2\Phi' + (1-r\Phi')^2f'\Big)f' -f''- 2(1-r\Phi') \left(-2\Phi'^2+\Phi''-3r\Phi'\Phi'' \right)f^2\nonumber\\  
&&\qquad+ \Big((1-r\Phi')^2f'' + 2\Phi'' -2 (-1+r\Phi' )f'\left(-3\Phi' +2r\Phi'^2-r \Phi''\right)  \Big)f \Bigg]=0,\label{A2}\\
&&\exp[3\Phi]\Bigg[ 1 -2r^2\mathcal{L}(\mathcal{F}) -(r+2\alpha\Phi')f' + \Bigg(-1+\alpha\Phi' \Big(-2(1+f)\Phi' +r^2f\Phi'^3\nonumber\\
&&\qquad\qquad +2\Big(3+r\Phi'(-3+r\Phi' ) \Big)f'\Big) + 4\alpha\Big(-1+(-1+r\Phi' )^2f\Big)\Phi''\Bigg)f \Bigg]=0.\label{A3}
\end{eqnarray}
\end{linenomath}
\end{adjustwidth}

Solving Equation~(\ref{A1}), leads to the solution for the scalar field as follows:
\begin{equation}
\Phi(r)=\log\Big[\frac{r}{L}\Big] + \log[\cosh(\xi)-\sinh(\xi)], \quad \xi(r)=\int_{1}^{r}\frac{du}{u\sqrt{f(u)}},\label{A4}
\end{equation}
where $L$ is an integration constant. For the scalar field (\ref{A4}), the dynamical equation for $\Phi(r)$ in (\ref{A2}) is automatically satisfied, and using the $4D$ NED Lagrangian density $\mathcal{L}(\mathcal{F})$ from Equation~(\ref{ned}), Equation~(\ref{A3}) yield the solutions for metric function $f(r)$ as
\begin{equation}
f_{\pm}(r)=1+\frac{r^2}{2\alpha}\left(1\pm\sqrt{1+\frac{8M\alpha}{(r^2+g^2)^{3/2}}}\right).\label{fr1}
\end{equation}

Though this approach of $4D$ regularization of EGB gravity is noteworthily different in spirit from the one proposed by Glavan and Lin \cite{Glavan:2019inb}, interestingly, the two theories yield exactly the same static spherically symmetric black hole solutions. Nevertheless, a larger class of black hole solutions may exist in the $4D$ effective scalar-tensor gravity theory followed by the Kaluza--Klein approach \cite{Hennigar:2020lsl,Lu:2020iav,Ma:2020ufk}. We would not however like to address the issues of a consistent $4D$ EGB theory, and so on, rather, our concern is for the regular black hole solution, which eventually turns out the same in both theories.

{Note added in proof:} After this work was completed, we learned of a similar work by Singh and Siwach \cite{Singh:2020xju}, which appeared in arXiv a couple of days before.

\begin{adjustwidth}{-\extralength}{0cm}

\reftitle{References}

\end{adjustwidth}
\end{document}